\documentclass[twocolumn,prb]{revtex4}

\usepackage{graphicx}

\begin{document}


\title{Alternating domains with uniaxial and biaxial magnetic anisotropy in epitaxial Fe films on BaTiO$_3$} 



\author{Tuomas H. E. Lahtinen,$^{1}$ Yasuhiro Shirahata,$^{2}$ Lide Yao,$^{1}$ K\'{e}vin J. A. Franke,$^{1}$ Gorige Venkataiah,$^{2}$ Tomoyasu Taniyama,$^{2}$ and Sebastiaan van Dijken$^{1}$}
\email{sebastiaan.van.dijken@aalto.fi}
\affiliation{$^{1}$NanoSpin, Department of Applied Physics, Aalto University School of Science, P.O. Box 15100, FI-00076 Aalto, Finland.} 
\affiliation{$^{2}$Materials and Structures Laboratory, Tokyo Institute of Technology, 4259 Nagatsuta, Midori-ku, Yokohama, Japan.}


\date{\today}

\begin{abstract}
We report on domain formation and magnetization reversal in epitaxial Fe films on ferroelectric BaTiO$_3$ substrates with ferroelastic $a-c$ stripe domains. The Fe films exhibit biaxial magnetic anisotropy on top of $c$ domains with out-of-plane polarization, whereas the in-plane lattice elongation of $a$ domains induces uniaxial magnetoelastic anisotropy via inverse magnetostriction. The strong modulation of magnetic anisotropy symmetry results in full imprinting of the $a-c$ domain pattern in the Fe films. Exchange and magnetostatic interactions between neighboring magnetic stripes further influence magnetization reversal and pattern formation within the $a$ and $c$ domains.           
\end{abstract}


\maketitle 

Ferromagnetic pattern formation via efficient coupling to ferroelectric domain structures has recently been demonstrated. \cite{2008NatMa...7..478C,2009PhRvL.103y7601L,2011PhRvL.107u7202H,ADMA:ADMA201100426,2011ITM....47.3768L,SciRep} Direct correlations between ferromagnetic and ferroelectric domains and its persistence during ferroelectric polarization reversal open up promising ways for electric-field control of local magnetic switching \cite{2008NatMa...7..478C,2009PhRvL.103y7601L,2011PhRvL.107u7202H,ADMA:ADMA201100426,2011ITM....47.3768L} and the motion of magnetic domain walls.\cite{SciRep} In systems based on interlayer strain transfer, the ferroelastic domain structure of a ferroelectric material induces local magnetoelastic anisotropies in a ferromagnetic film via inverse magnetostriction. Within the ferromagnetic sub-system, the magnetoelastic anisotropy competes with intrinsic magnetic properties including magnetocrystalline anisotropy and exchange and magnetostatic interactions between domains. Consequently, the evolution of the magnetic microstructure in an applied magnetic or electric field depends critically on the two ferroic materials, the ferromagnetic layer thickness, and the ferroelastic domain size.

In a previous study, full pattern transfer from ferroelectric BaTiO$_3$ substrates with alternating ferroelastic $a_1-a_2$ domains to polycrystalline Co$_{60}$Fe$_{40}$ thin films was analyzed. \cite{ADMA:ADMA201100426,2011ITM....47.3768L,SciRep} The strain-induced uniaxial magnetoelastic anisotropy axis of this system rotates by 90$^\circ$ at domain boundaries and this fully dominates the local magnetic properties since the magnetocrystalline anisotropy of Co$_{60}$Fe$_{40}$ is negligibly small. Here, we report for the first time on full imprinting of ferroelectric BaTiO$_3$ $a-c$ domain patterns into epitaxial Fe films as schematically illustrated in Fig. 1(a). Magneto-optical Kerr effect (MOKE) microscopy measurements indicate that the magnetic anisotropy is laterally modulated by the alternating in-plane structural symmetry of the BaTiO$_3$ lattice. Moreover, it is shown that magnetic switching in neighboring $a$ and $c$ domains is strongly coupled via exchange and magnetostatic interactions. The demonstrated ability to initialize a rich variety of micromagnetic configurations in Fe/BaTiO$_3$ supports the design of electric-field controlled magnetic structures including magnonic crystals and spintronic devices.        

\begin{figure} [b]
\includegraphics{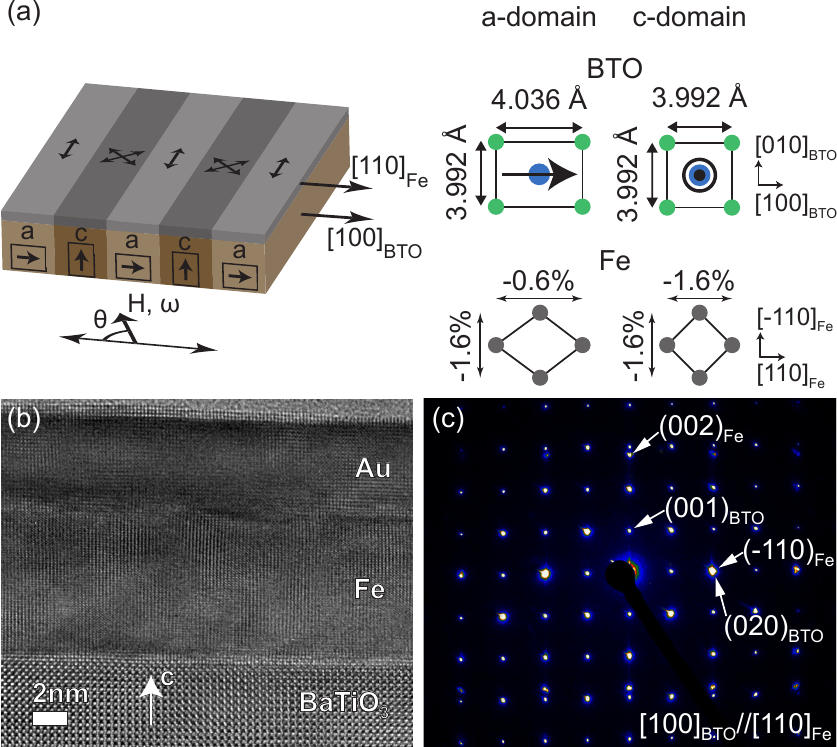}
\caption{\label{figure} (a) Schematic illustration of the domain configuration in the Fe/BaTiO$_3$ system. The arrows in the BaTiO$_3$ substrate indicate the direction of ferroelectric polarization and the double-headed arrows in the Fe film represent the magnetic easy axes. The schematics on the right illustrate the in-plane lattice structure and polarization direction of the BaTiO$_3$ substrate and the orientation of the Fe film in the $a$ and $c$ domains. The room-temperature lattice parameters of the BaTiO$_3$ substrate and the in-plane strains of the Fe film are indicated. The strains are calculated relative to the Fe bulk lattice parameter on the basis of full strain transfer between substrate and film. A cross-sectional TEM image and a selected area electron diffraction (SAED) pattern of a 10 nm thick epitaxial Fe film on BaTiO$_3$ (001) are shown in (b) and (c).}
\end{figure}

\begin{figure*} [t]
\includegraphics{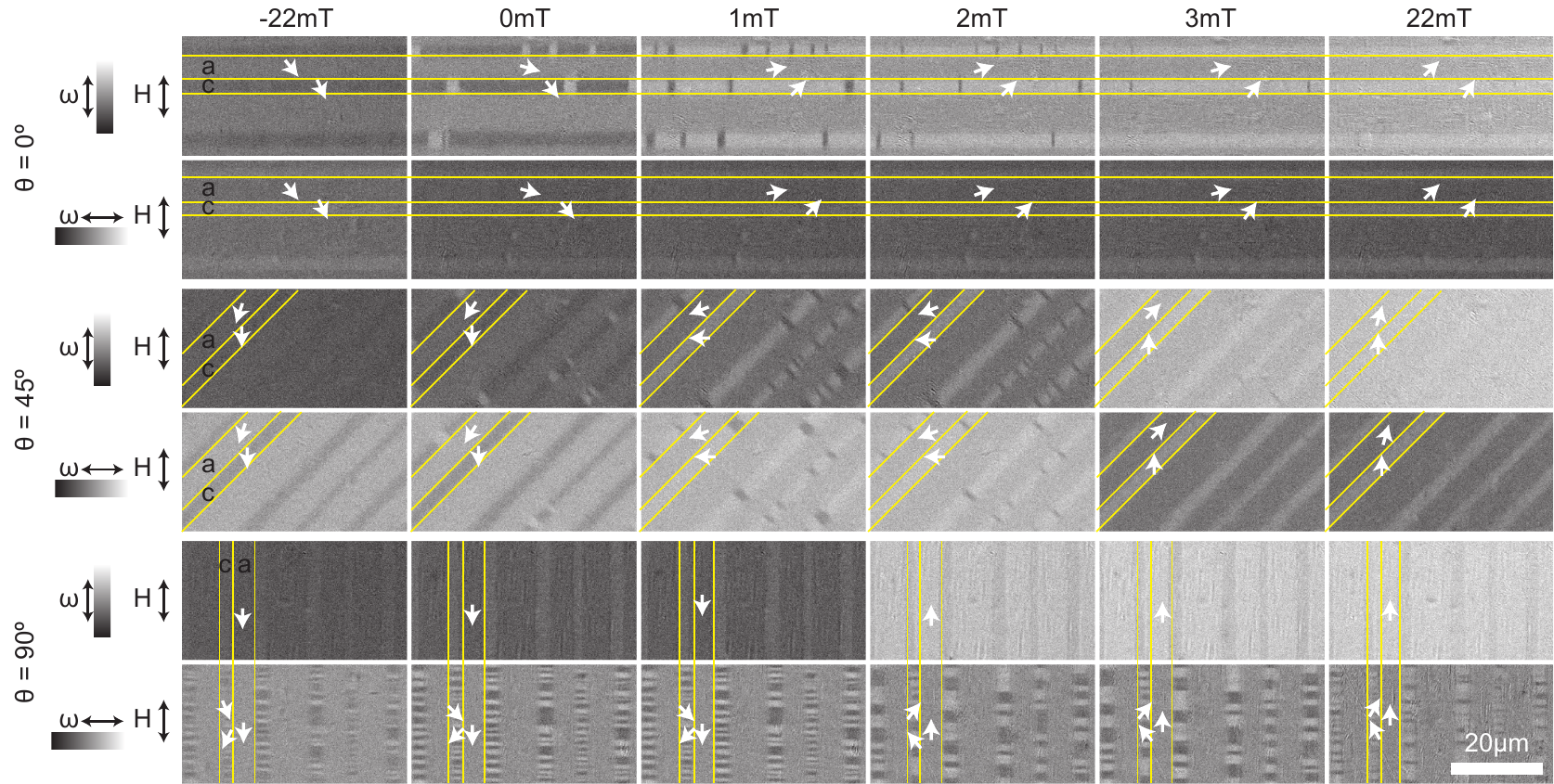}
\caption{\label{Figure2} Kerr microscopy images of a 20 nm thick epitaxial Fe film on BaTiO$_3$ for different magnetic field angles ($\theta$) and field strength. The direction of magnetic field was altered by sample rotation in the Kerr microscope while the electromagnet was fixed. Negative fields point down and positive fields point up in the above images. The white arrows indicate the direction of magnetization within selected $a$ and $c$ domains as inferred from a set of two images with the magneto-optical contrast axis ($\omega$) along the vertical and horizontal direction, respectively.}
\end{figure*}

In the experiments, 10 nm and 20 nm thick Fe films with a 5 nm Au capping layer were grown onto single-crystal BaTiO$_3$ substrates using molecular beam epitaxy \cite{2011ApPhL..99j2506V,2011ApPhL..99b2501S}. Film growth at 300$^\circ$C in an ultrahigh vacuum chamber resulted in epitaxial Fe with an in-plane Fe [110] $\parallel$ BaTiO$_3$ [100] alignment as confirmed by transmission electron microscopy (Fig. 1(b) and (c)). The ferroelastic domain pattern of the BaTiO$_3$ substrates consisted of alternating $a$ and $c$ stripe domains with an average width of 6 $\mu$m and 3 $\mu$m at room temperature. This regular domain structure was formed upon sample cooling through the ferroelectric Curie temperature ($T_C=120^\circ$C) after Fe film growth. At $T_C$, the structure of BaTiO$_3$ changes from cubic to tetragonal. The ferroelastic $a$ and $c$ domains in tetragonal BaTiO$_3$ impose different local strains on the epitaxial Fe films. For full strain transfer, the in-plane lattice of the Fe films is compressed by $-1.6\%$ and $-0.6\%$ with respect to that of the bcc Fe bulk structure ($a_{Fe} = 2.870$ \AA) as illustrated in Fig. 1(a). On top of the ferroelastic $c$ domains, the Fe lattice remains cubic. However, due to different shear strains in the [110]$_{\textrm{Fe}}$ and [-110]$_{\textrm{Fe}}$ directions, the in-plane structure of the Fe films changes into a diamond shape on top of the BaTiO$_3$ $a$ domains. MOKE microscopy was used to analyze magnetization reversal in the $a$ and $c$ domains as a function of in-plane magnetic field. Images were recorded with two orthogonal magneto-optical contrast axes ($\omega$) by rotation of the optical plane of incidence using an adjustable diaphragm. From these two images, the local magnetization direction was extracted. The MOKE microscope setup was also used to measure magnetic hysteresis curves on single $a$ and $c$ stripe domains for different in-plane magnetic field angles ($\theta$).

Figure 2 shows MOKE microscopy images of a 20 nm thick Fe film on top of BaTiO$_3$ for three magnetic field directions. The regular magnetic stripes directly correlate with the ferroelastic $a-c$ pattern of the BaTiO$_3$ substrate. The magnetic domain walls are strongly pinned onto the ferroelectric domain boundaries by abrupt changes in local magnetic anisotropy. \cite{SciRep,PhysRevB.85.094423} As a result, the overall stripe pattern does not alter in an applied magnetic field (until the film is saturated) and magnetic switching in the $a$ and $c$ domains differs considerably. For a magnetic field angle of  $\theta$ = 0$^\circ$ ($H$ perpendicular to the domains), the magnetization of the $a$ domains rotates coherently as indicated by the white arrows in the images on the top rows of Fig. 2. On the other hand, abrupt switching is observed for $\theta$ = 90$^\circ$ ($H$ parallel to the domains). In the images on the bottom rows of Fig. 2, this switching event can be clearly seen at 2 mT. The difference in magnetization reversal mechanism reflects the uniaxial magnetic symmetry of the $a$ domains, which is also confirmed by the local magnetic hysteresis curves of Fig. 3 and the polar plot of the remnant magnetization in Fig. 4(a). The magnetic easy axis of the $a$ domains in the epitaxial Fe films is oriented parallel to the magnetic domains ([-110]$_{\textrm{Fe}}$ direction) and, thus, perpendicular to the in-plane lattice elongation and polarization direction of the ferroelectric BaTiO$_3$ substrate. This qualitatively agrees with the symmetry of magnetoelastic anisotropy in Fe/BaTiO$_3$ as illustrated in the next paragraph. 

\begin{figure} [t] 
\includegraphics{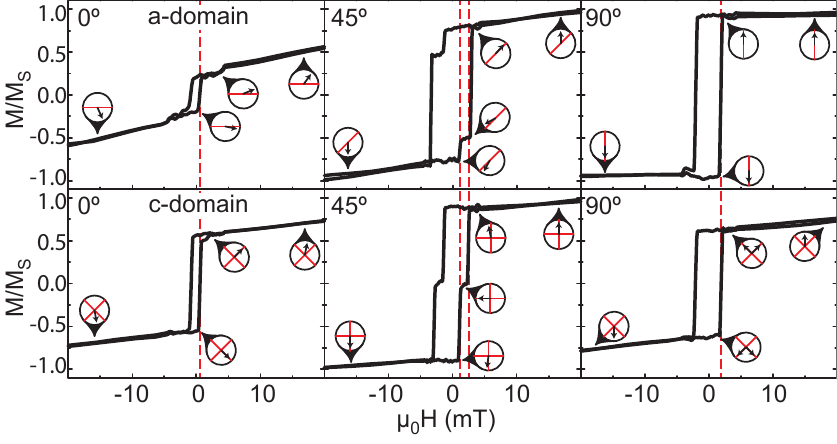}
\caption{\label{Figure3} Magnetic hysteresis curves of single $a$ (top row) and $c$ (bottom row) domains measured at different magnetic field angles. The red lines and arrows in the circular symbols indicate the easy magnetization axes and the direction of magnetization during different stages of the reversal process. The vertical red dashed lines illustrate simultaneous magnetic switching in the $a$ and $c$ domains.}
\end{figure}

The magnetoelastic anisotropy energy density ($K_{me}$) of Fe films on top of BaTiO$_3$ can be written as \cite{Handley}
\begin{equation}
	K_{me} = B_1(e_{xx}\alpha_x^2+e_{yy}\alpha_y^2) + B_2e_{xy}\alpha_x\alpha_y 
\end{equation}
Here, $B_1$ and $B_2$ are the magnetoelastic coupling coefficients, $e_{xx}$ and $e_{yy}$ are the strains along the cubic axes ([100]$_{\textrm{Fe}}$ and [010]$_{\textrm{Fe}}$), $e_{xy}$ is the shear strain along the diagonal [110]$_{\textrm{Fe}}$ direction, and $\alpha_x$ and $\alpha_y$ are the directional cosines with respect to [100]$_{\textrm{Fe}}$ and [010]$_{\textrm{Fe}}$. On top of the ferroelastic $a$ domains, the in-plane Fe lattice exhibits a diamond shape due to a shear strain along [110]$_{\textrm{Fe}}$. Relative to the [-110]$_{\textrm{Fe}}$ direction, the magnitude of this strain amounts to $+1.0\%$ for full strain transfer. The change in magnetoelastic anisotropy that this strain effect induces is given by the second term of Eq. 1. Using $B_2=7.83\times10^6$ J/m$^3$, \cite{1999RPPh...62..809S} $e_{xy}=1.0\%$, and $\alpha_x=\alpha_y=1/\sqrt2$, this calculation yields $\Delta$$K_{me}=3.9\times10^4$ J/m$^3$. Since $\Delta$$K_{me}$ is positive, the [110]$_{\textrm{Fe}}$ direction is magnetically hard. Consequently, the uniaxial magnetic easy axis is oriented along the orthogonal [-110]$_{\textrm{Fe}}$ direction, i.e. parallel to the magnetic domains and perpendicular to the BaTiO$_3$ ferroelectric polarization. The strength of the magnetoelastic anisotropy as determined from the slope and saturation field of hard-axis hysteresis curves is $K_{me}=5\pm2\times10^4$ J/m$^3$. The close agreement between the calculated and experimentally measured uniaxial magnetoelastic anistropy indicates that the ferroelastic strain of the BaTiO$_3$ substrates is fully transferred to the epitaxial Fe films when the samples are cooled after MBE growth. 

\begin{figure} [t]
\includegraphics{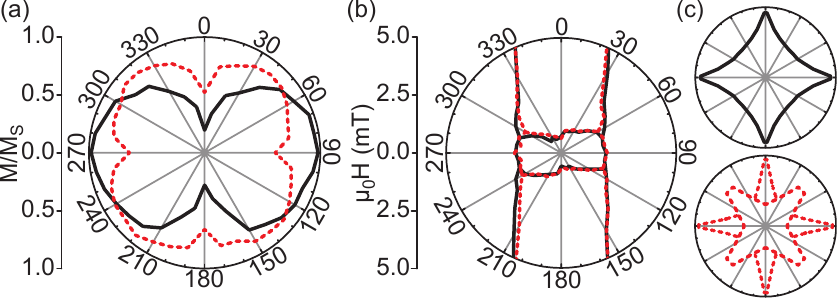}
\caption{\label{Figure4} Experimental polar plots of (a) the remnant magnetization and (b) the magnetic switching fields of the $a$ (black solid lines) and $c$ (red dashed lines) magnetic stripe domains. (c) Polar plot of the switching fields for isolated uniaxial $a$ and biaxial $c$ domains as determined by micromagnetic simulations. For both domains, an anisotropy constant of $K=5\times10^4$ J/m$^3$ was used in the simulations.}
\end{figure}

Magnetization reversal in the $c$ domains of the Fe film is strikingly different. The polar plot of the remnant magnetization in Fig. 4(a) clearly illustrates that the magnetic anisotropy of the $c$ domains exhibits fourfold symmetry. The easy magnetization axes are aligned along [100]$_{\textrm{Fe}}$ and [010]$_{\textrm{Fe}}$, in agreement with the magnetocrystalline anisotropy of bulk Fe. Moreover, the in-plane lattice of BaTiO$_3$ remains cubic when $c$ domains form during sample cooling through $T_C$. The lattice misfit between Fe ($a_{Fe} = 2.870$ \AA) and BaTiO$_3$ ($a_{BaTiO_3} = 3.992$ \AA) at room temperature therefore induces a biaxial compressive strain of $-1.6\%$ along the cubic axes. From the first term in Eq. 1 and $e_{xx}=e_{yy}-1.6\%$, $\alpha_x=1$, $\alpha_y=0$, and $B_1=-3.43\times10^6$ J/m$^3$ \cite{1999RPPh...62..809S}, the magnetoelastic anisotropy along [100]$_{\textrm{Fe}}$ can be estimated. This gives $K_{me}=-5.5\times10^4$ J/m$^3$. The same result is obtained for [010]$_{\textrm{Fe}}$ with $\alpha_x=0$ and $\alpha_y=1$. Thus, the biaxial compressive strain along [100]$_{\textrm{Fe}}$ and [010]$_{\textrm{Fe}}$ lowers the magnetoelastic anisotropy energy along these directions, which agrees with the experimentally observed orientation of the easy magnetization axes in the $c$ domains (Fig. 4).       

Besides the strong lateral modulation of magnetic anisotropy, local magnetization reversal in Fe/BaTiO$_3$ also depends on the coupling between neighboring magnetic stripe domains. This is clearly illustrated by the double switching events in the hysteresis curves of the $a$ and $c$ domains and the close agreement in switching fields (Fig. 3). Double switching has been observed for epitaxial Fe wires with biaxial anisotropy and field angles near the magnetic easy axes, \cite{PhysRevB.56.5443} but the occurrence of two consecutive reversal events in the uniaxial $a$ domains can only be rationalized when magnetic interactions between neighboring domains are taken into account. To illustrate the coupling between the Fe stripe domains and the nature of the magnetic interactions, the hysteresis curves of Fig. 3 are analyzed in detail. 

For $\theta$ = 0$^\circ$, the magnetic field is aligned along the uniaxial magnetic hard axis of the $a$ domains. However, in addition to coherent magnetization reversal, an abrupt magnetic switch is observed in the hysteresis curve. This switch is induced by 90$^\circ$ magnetization rotation in the neighboring $c$ domains at 0.6 mT. Because ferromagnetic coupling between the $a$ and $c$ domains reduces the magnetization angle just prior and after this reversal event, the magnetization of the $a$ domains instantaneously rotates when the $c$ domains switch. The induced magnetization rotation in the $a$ domains, as estimated from local MOKE hysteresis curves, is about 20$^\circ$. Since exchange and magnetostatic interactions both reduce the magnetization angle between domains in this configuration, it is not possible to distinguish between both magnetic effects. 

For $\theta$ = 90$^\circ$, the magnetization of the $a$ domains is aligned  along the uniaxial magnetic easy axis throughout the reversal process. In this case, one single switching event is observed at 2.0 mT in both stripe domains. The increase of switching field compared to $\theta$ = 0$^\circ$ indicates that ferromagnetic coupling between neighboring domains delays magnetization reversal in the biaxial $c$ domains until the $a$ domains switch by 180$^\circ$. This domain coupling effect can be rationalized by considering the magnetic energy of the Fe/BaTiO$_3$ system. If the $c$ domains would switch prior to the $a$ domains, the magnetization angle between both domains would increase from about 45$^\circ$ to 135$^\circ$ (see circular symbols in Fig. 3). This would enhance both the exchange and magnetostatic energy. Simultaneous switching in both domains is therefore energetically more favorable because it limits the maximum magnetization angle between neighboring domains to 45$^\circ$ throughout the reversal process. Since the magnetization of the $a$ domains is parallel to the stripe pattern, their stray field hardly influences the magnetization direction in the $c$ domains. Hence, ferromagnetic coupling between both stripe domains is dominated by short-range exchange interactions. 

Finally, for $\theta$ = 45$^\circ$, the magnetization of the $c$ domains reverses by two separate 90$^\circ$ switches. In this case, both coupling effects are observed. At small magnetic field, magnetic switching in the biaxial $c$ domains induces a small abrupt magnetization rotation in the uniaxial $a$ domains. The second switch in the $c$ domains is triggered by 180$^\circ$ magnetization reversal in the $a$ domains. Both exchange and magnetostatic interactions might contribute to domain coupling in this case.  

The polar plot of Fig. 4(b) summarizes the switching fields of the $a$ and $c$ domains as a function of magnetic field angle. Strong coupling between neighboring stripe domains results in identical switching fields for both domains irrespective of field direction. The vertical lines represent magnetic switching events that are triggered by uniaxial $a$ domains. The shape of these curves deviates significantly from the Stoner-Wohlfarth astroid,\cite{1948RSPTA.240..599S} which is obtained when the switching fields of isolated uniaxial domains are simulated (Fig. 4(c)). In the Stoner-Wohlfarth case, a maximum switching field is obtained when the field is applied along the uniaxial magnetic easy axis ($\theta$ = 90$^\circ$). In our Fe/BaTiO$_3$ samples, strong coupling between neighboring stripe domains rotates the magnetization of the $a$ domains away from its magnetic easy axis during magnetization reversal and this drastically reduces the switching field for $\theta$ = 90$^\circ$. Similarly, the horizontal lines in Fig. 4(b) indicate magnetic switching events that are induced by biaxial $c$ domains. Again, the shape of these experimental curves differs completely from the simulated polar plot of isolated domains with biaxial anisotropy (Fig. 4(c)). The measured polar plot of the switching fields in Fe/BaTiO$_3$ is unique. Both the vertical and horizontal curves have no analogue in systems with uniform magnetic anisotropy. We also note that our microscopic observation of a small intrinsic $c$-domain switching field agrees with previous macroscopic studies demonstrating a reduction of magnetic switching field after the application of an out-of-plane electric field. \cite{2011ApPhL..99j2506V,2011ApPhL..98i2505B}  

Finally, we show that efficient coupling between $a$ and $c$ domains influences the magnetic pattern within $a$ and $c$ stripe domains of the Fe films. If both domains would switch independently, the evolution of the magnetic microstructure of the biaxial $c$ domains would be similar for $\theta$ = 0$^\circ$ and $\theta$ = 90$^\circ$. The MOKE microscopy images of Fig. 2 clearly show that this is not the case. For $\theta$ = 0$^\circ$, the magnetization of the $a$ domains coherently rotates clockwise and, due to ferromagnetic interactions between domains, this forces the magnetization of the $c$ domains to reverse in the same direction. In small magnetic field, 90$^\circ$ domains nucleate within the $c$ stripes. These domains are highly mobile and their size increases rapidly with applied field strength. As a result, the magnetic uniformity of the $c$ domains is re-established at a field of about 3 mT. For $\theta$ = 90$^\circ$, the magnetization of the $a$ domains remains fixed along its uniaxial magnetic easy axis. In this case, exchange interactions with neighboring $c$ domains do not favor a particular reversal direction. The $c$ domains therefore split up into a regular stripe pattern by alternating clockwise and anti-clockwise magnetization rotation. Due to negligible domain wall motion within the $c$ domains under these conditions, the stripe pattern is robust until it abruptly changes during simultaneous magnetic switching in both domains. In this case, a large magnetic field is required to fully remove the stripe pattern from the $c$ domains.           

In summary, the results of this paper clearly demonstrate that magnetic domains can be imprinted into continuous magnetic films via efficient coupling to ferroelastic domains of a ferroelectric material. The magnetization reversal mechanism in such domains depends on the symmetry of the strain-induced magnetoelastic anisotropy and the applied field angle. Moreover, ferromagnetic coupling between neighboring domains strongly influences local magnetic switching events. Pattern formation in epitaxial Fe films on ferroelectric $a-c$ domains of BaTiO$_3$ is particularly rich. In this system, magnetic stripe domains with biaxial and uniaxial anisotropy alternate. Moreover, depending on the direction of the applied magnetic field, highly mobile or robust magnetic patterns form within the $c$ domains, a property that could be exploited in magnetic devices. 



%
%

%

\begin{acknowledgments}
This work was supported by the Academy of Finland (Grant Nos. 127731 and 260361), the European Research Council (ERC-2012-StG 307502-E-CONTROL), the Industrial Technology Research Grant Program in 2009 from NEDO of Japan, JSPS KAKENHI (Grant No. 2200077), the Advanced Materials Development and Integration of Novel Structured Metallic and Inorganic Materials Project of MEXT, and the Collaborative Research Project of the
Materials and Structures Laboratory, Tokyo Institute of Technology. T.H.E.L. is supported by the National Doctoral Program in Materials Physics and K.J.A.F. acknowledges support from the Finnish Doctoral Program in Computational Sciences. 
\end{acknowledgments}


\end{document}